\documentclass{article}
\usepackage{epsfig,amsmath,amssymb}

\DeclareMathSymbol{\varGamma}{\mathord}{letters}{"00}

\newcommand{\vk}{\vec{k}}

\begin{document}
\title{RADIATIVE DECAYS OF HADRONIC MOLECULES}
\author{A.V. Nefediev\\
{\em Institute of Theoretical and Experimental Physics,} \\
{\em 117218, B.Cheremushkinskaya 25, Moscow, Russia}}
\date{}
\maketitle
\baselineskip=11.6pt
\begin{abstract}
It is argued that radiative decays of scalars $a_0/f_0(980)$ can serve as a
decisive tool in establishing the nature of the latter. In particular,
predictions for the widths of the radiative decays $S\to \gamma V$
($S=a_0/f_0(980)$, $V=\omega/\rho/\gamma$) are given in the framework of the
molecule model of the scalars. Finite--range corrections are discussed in
detail for the two-gamma decays of hadronic molecules, with a special
attention payed to the interplay of various scales involved in the
problem and to the gauge invariance of the amplitude. The results are
applied to the two-photon decay of the $f_0(980)$, and the existing
experimental data on this decay are argued to support the molecule
assignment for the scalar $f_0(980)$.
\end{abstract}
\baselineskip=14pt

The problem of the structure of light scalar mesons is of a fundamental 
importance for understanding the properties of the entire scalar sector, 
that is, the sector of states with the quantum numbers of the vacuum, 
including purely gluonic excitations. In particular, the identification of 
the $a_0(980)$ and $f_0(980)$ mesons, together with the experimental studies of the 
lightest scalars ($\sigma$ and $\kappa$), will allow one to establish the 
structure of multiplets of scalars and to find the signature of the scalar 
glueball in the spectrum of physical states. 
There are several models for the $a_0(980)$ and $f_0(980)$. The latter can be considered as
${}^3P_0$ quark--antiquark states \cite{qq} strongly coupled to the mesonic continuum and thus strongly distorted with the 
unitarisation process. However, due to the proximity of the $K\bar{K}$ 
threshold, it is natural to assume a considerable admixture of the 
four--quark component in the wave functions of these mesons, 
either as a compact four--quark with hidden strangeness \cite{Jaffe,4q}, or as a $K\bar{K}$ 
molecule. These might be t-channel exchanges to be responsible for the 
formation of such a molecule \cite{WeiIs,Jue,Oset,Markushin}. 
It is therefore important to establish a test which would allow one to distinguish between these models and thus to
reveal the actual nature of these scalars (in particular, efficient methods to discriminate between the molecule and compact states are
strongly needed --- for the recent progress see \cite{Ch}).
Since years, radiative decays of the $\phi(1020)$, $\phi\to\gamma S$, have been considered as such an
experimental tool \cite{AI}. Indeed, these decays point to a large $K\bar{K}$ component in the scalars wave function \cite{SND,CMD,KLOE}. 
Still a number of
shortcomings of this approach should be mentioned. First of all, the radiative decays of the $\phi$ do not allow one to probe the nonstrange
component of the scalars, and the contribution of the quark loops is strongly suppressed as compared to the contribution of the meson loops.
Finally, the phasespace available in the final state of these decays is limited to a large extend. In the meantime, another class of radiative
decays involving scalars is known --- the radiative decays of the scalars themselves: $S\to\gamma V$, where the vector in the final state is
either massive ($\rho$ or $\omega$) or massless, that is one deals with a two--photon decay in the latter case. 
Whatever model of scalars is used, gauge invariance imposes strong constraints on the decay amplitude:
\begin{equation}
iW^{\mu\nu}=M(a,b)[P_V^\mu P_\gamma^\nu-g^{\mu\nu}(P_VP_\gamma)],\quad a=\frac{m^2_V}{m^2},\quad b=\frac{m^2_S}{m^2},
\label{W1}
\end{equation}
where $m$ is the kaon mass and $P_{V,\gamma}$ are the four--momenta of the vector particles. Below we shall evaluate the widths of the radiative decays
involving the scalars $a_0/f_0(980)$ in the molecule assignment for the latter. 

\begin{figure}[t]
\begin{center}
\begin{tabular}{ccc}
\epsfig{file=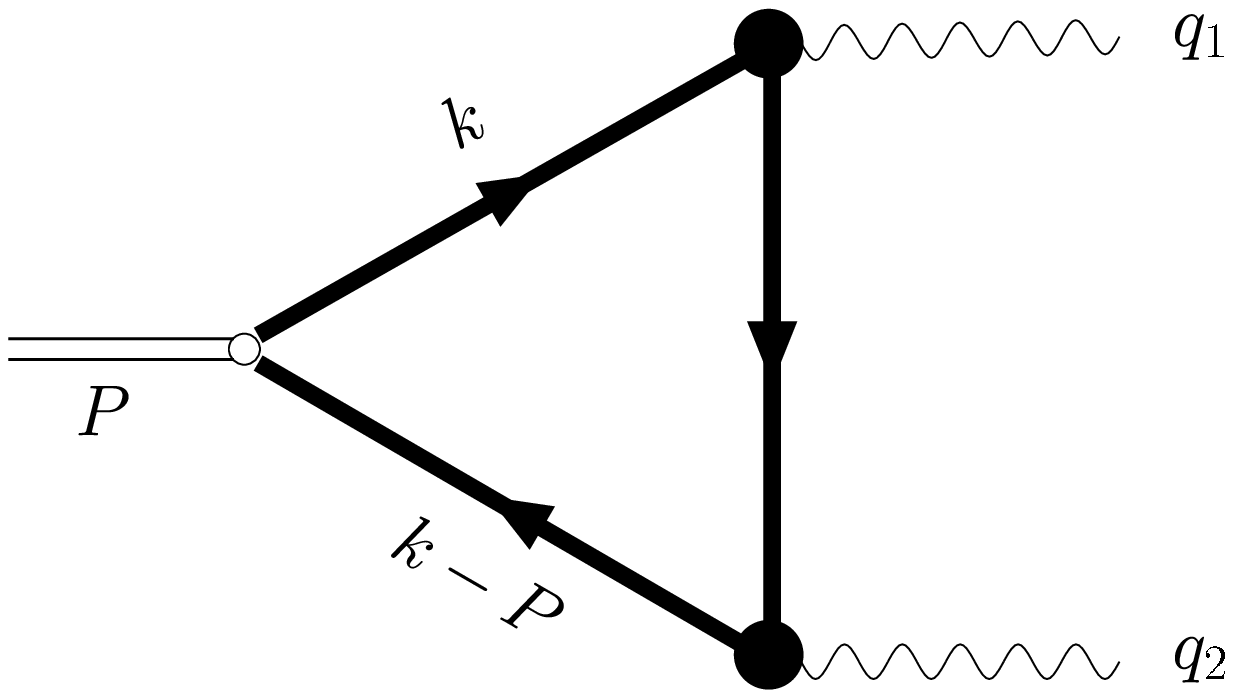,width=3.4cm}&\epsfig{file=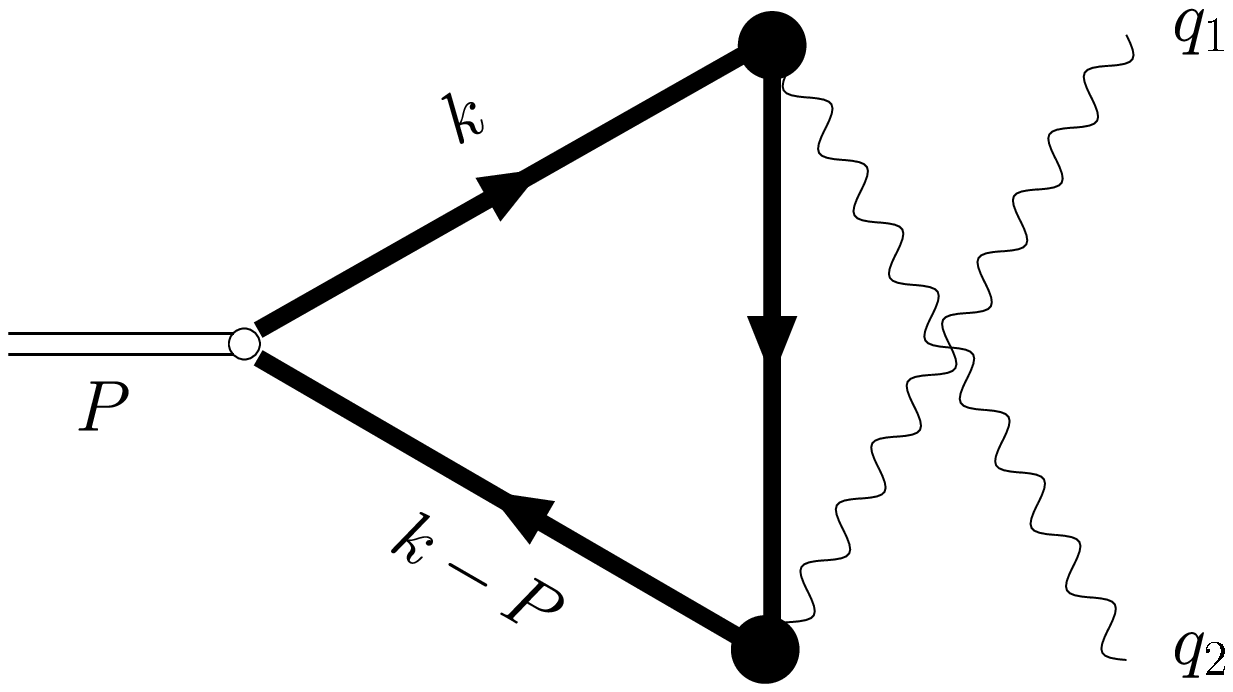,width=3.4cm}&\raisebox{3mm}{\epsfig{file=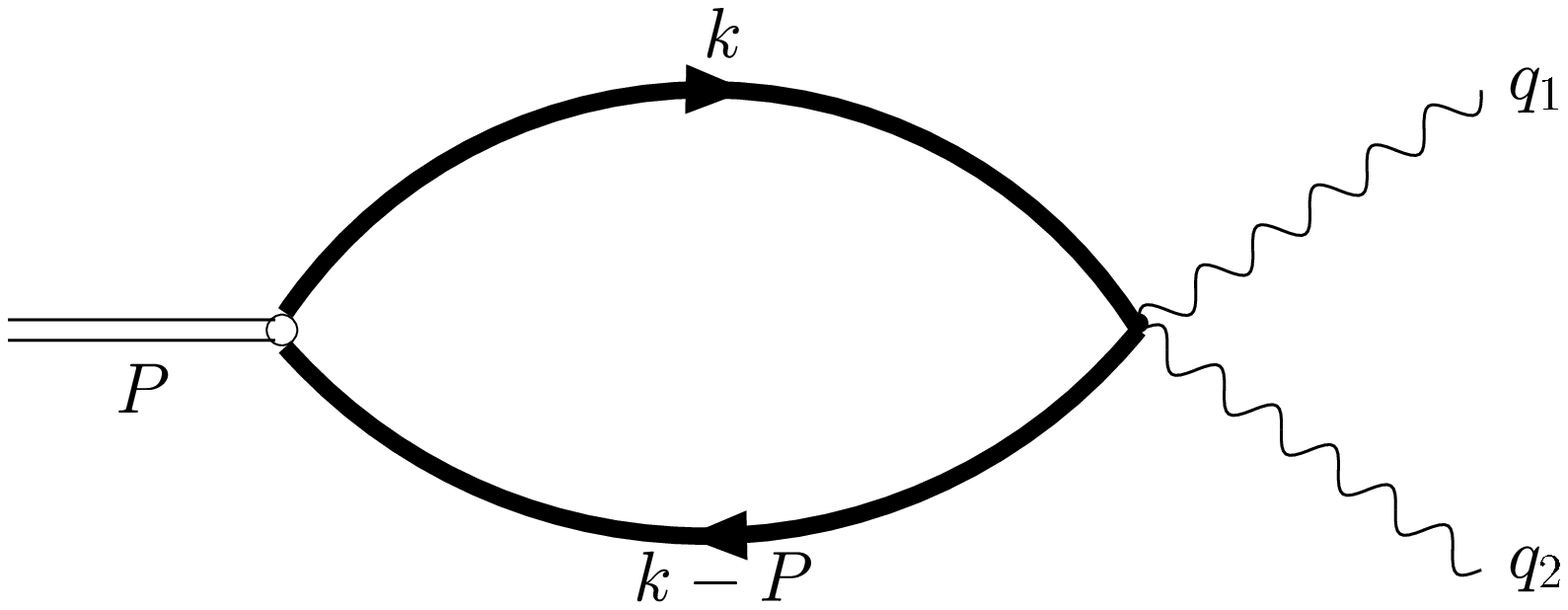,width=3.8cm}}\\
(a)&(b)&(c)\\
\end{tabular}
\end{center}
\caption{\it Diagrams contributing to the scalar decay amplitude.}\label{3d}
\end{figure}

First of all, it is important to notice that there are three scales in the problem under consideration, which are (i) the binding force scale
$\beta\simeq m_\rho\approx 800$ MeV, (ii) the kaon mass $m$, and (iii) the binding energy $\varepsilon$, and the hierarchy of these scales is
$\varepsilon\ll m\lesssim\beta$. The last inequality suggests that it is natural to start from the point-like limit of $\beta\to\infty$ and
to include finite--range corrections (in the form a $1/\beta$ expansion) afterwards. Thus we stick to the point-like limit.
The first ingredient one needs to know is the coupling of the loosely bound molecule state to the $K\bar{K}$ pair, which 
reads \cite{mol0}:
\begin{equation}
\frac{g_S^2}{4\pi}=32m\sqrt{m\varepsilon}\approx 1.12\;{\rm GeV}^2,
\label{gS0}
\end{equation}
where $m=495$ MeV and the molecule binding energy is taken to be $\varepsilon=10$ MeV. In addition, the $\phi K\bar{K}$ and the
$VK\bar{K}$ coupling constants can be evaluated using the total width of the $\phi$ and the $\rho\pi\pi$ constant under the assumption of 
the SU(3) invariance. One arrives then at $g_\phi=4.72\quad g_V=2.13$. It is straightforward then to arrive at the predictions of the
point-like model for the radiative decays involving scalars. We give these predictions in tab.\ref{t1}.
For illustrative purposes and for future references, let us quote the formula for the two-photon decay width of a point-like scalar (see fig.\ref{3d} for the diagrams
contributing to this decay):
\begin{equation}
\Gamma(S\to\gamma\gamma)=\frac12\left(\frac{\alpha}{\pi}\right)^2\sqrt{m\varepsilon}\left(\frac{2m}{m_S}\right)
\left[\left(\frac{2m}{m_S}\right)^2\arcsin^2\left(\vphantom{\frac{2m}{m_S}}\frac{m_S}{2m}\right)-1\right]^2,
\label{beauty}
\end{equation}
where $m_S=2m-\varepsilon$. 

\begin{table}[t]
\centering
\caption{\it The widths (in keV) of the radiative decays involving scalars; 
$\theta$ is the (small) $\phi-\omega$ mixing angle.}\label{t1}
\vskip 0.1 in
\begin{tabular}{|l|c|c|c|}
\hline
&Quark--antiquark&Molecule&Data (PDG)\\
\hline
\hline
$\phi\to\gamma a_0$&$0.37\sin^2\theta$&0.6&$0.32\pm0.02$\\
\hline
$\phi\to\gamma f_0(\bar{n}n)/f_0(\bar{s}s)$&$0.04\sin^2\theta/0.18$&0.6&$0.47\pm0.03$\\
\hline
$a_0\to\gamma\gamma$&$\sim 1$&0.22&$0.30\pm 0.10$\\
\hline
$f_0\to\gamma\gamma$&$\sim 1$&0.22&$0.29^{+0.07}_{-0.09}$\\
\hline
$a_0\gamma\omega/\rho$&125/14&3.4&\\
\cline{1-3}
$f_0(\bar{n}n)\gamma\rho/\omega$&125/14&3.4&pending\\
\cline{1-3}
$f_0(\bar{s}s)\gamma\rho/\omega$&$0/31\sin^2\theta$&3.4&\\
\hline
\hline
\end{tabular}
\end{table}

Notice that another approach to two--photon decays of molecules is known in the literature, namely the approach based on the 
formula $\Gamma(S\to\gamma\gamma)=\frac{\pi\alpha^2}{m^2}|\Psi (0)|^2$ which is written in analogy with that for the positronium two--photon
decay. Although this approach appears quite successful in QED, it has to fail in hadronic physics. First of all, the w.f. of the kaon
molecule is simply not known, so one has to rely on models. Moreover, since $\Psi(0)$ is very sensitive to the details of the bound--state formation,
the predictions of this approach may vary drastically (the predictions found in the literature vary by an order of magnitude, from 0.6 keV
in \cite{Barnes} to 6 keV in \cite{Krewald}). Furthermore, any attempt to evaluate corrections to this leading term results in either
gauge invariance or energy conservation law breaking. Indeed, the decay amplitude in this approach is usually given as an overlap integral
between the molecule w.f. and the amplitude of the process $K^+K^-\to\gamma\gamma$:
\begin{equation}
W\propto \int\frac{d^3k}{(2\pi)^3}\psi(\vk)\left[W(K^+(\vk)K^-(-\vk)\to\gamma\gamma)\right].
\label{M1}
\end{equation}
If the amplitude $W(K^+K^-\to\gamma\gamma)$ is taken off-shell, then it obviously fails to be gauge--invariant. On the contrary, for the on--shell
gauge--invariant amplitude $W(K^+K^-\to\gamma\gamma)$, kaons carry the energy $\sqrt{k^2+m^2}$, rather than $m_S/2$, so that energy conservation law is violated.

The last, but not the least, argument against using the $\Psi(0)$-based approach 
to hadronic processes is that the hierarchy of scales has to be different,
namely $\varepsilon\ll\beta\ll m$, in order to validate the given formula with $\Psi(0)$. 

Finally, we estimate the finite--range corrections to the point-like predictions quoted in tab.\ref{t1} --- we are interested in the
potentially large corrections of order $m^2/\beta^2$. If the vector in the final state is massive, then the photon in the final state is soft
($\omega\ll\beta$), and the kinematics of the loop becomes nonrelativistic \cite{Markushin,HKKN,mol0}. Inclusion of the finite-range
effects amounts to the substitution $g_S\to\varGamma(\vec{k})$, with a suitable form of the vertex $\varGamma(\vec{k})$. Gauge invariance
requires then that the extra momentum dependence coming from the vertex argument should be gauged, that is
\begin{equation}
\varGamma(\vec{k})\to\varGamma(\vec{k}+e\vec{A})\approx \varGamma(\vec{k})+e\vec{A}\frac{\partial\varGamma(\vec{k})}{\partial\vec{k}}
+{\cal O}\left(\frac{\omega^2}{\beta^2}\right),
\end{equation}
and the derivative term gives rise to an extra contact diagram with the photon emission from the scalar vertex. Gauge invariance is therefore 
preserved to order ${\cal O}\left(\omega/\beta\right)$. Notice however that one should be extremely careful when treating the loop integrals
entering the decay amplitude. Indeed, although the full amplitude is finite, every individual integral is divergent. If a cut-off is
introduced then to make them finite, gauge invariance may be badly broken and, as a result, wrong conclusions may be deduced (see, for 
example, \cite{Ast} and explanations in \cite{Ast2}). Finally one arrives at the conclusion that no corrections of order $m^2/\beta^2$ appear
for the point-like predictions \cite{Markushin,HKKN,mol0}. The same conclusion holds for the
two--photon decays of scalars, though it is not straightforward to arrive at this conclusion and one needs to develop a selfconsistent
gauge-invariant approach to this decay. It was suggested in \cite{2gam} to use an effective kaon interaction Lagrangian (for the
neutral-particle exchange, generalisation to the charged-particle exchange being trivial) written to order $1/\beta^2$:
\begin{equation}
L_{\rm int}=\frac12\lambda_1(\varphi^\dagger\varphi)^2+
\frac{\lambda_2}{2\beta^2}\left[\partial_\mu(\varphi^\dagger\varphi)\right]^2,
\label{Lint}
\end{equation}
with the coupling constants $\lambda_{1,2}$ being of the same order of magnitude. This Lagrangian is subject to renormalisation to order
$1/\beta^2$ --- see \cite{2gam} for the details. In the renormalised theory, the kaon propagator and the photon--emission vertex, which 
are the dressed quantities, built as solutions of the corresponding field theoretical equations, read:
\begin{equation}
S(p)=\frac{Z}{p^2-m^2},\quad v_\mu(p,q)=Z^{-1}(2p-q)_\mu+\ldots,
\label{ren1}
\end{equation}
where the ellipsis denotes terms which do not contribute to the decays under consideration. The renormalisation constants for the
kaon propagator and for the photon emission vertex coincide due to gauge invariance. The kaon mass $m$ is the renormalised physical mass.
Finally, the most important ingredient --- the scalar vertex beyond the point-like limit --- comes as a solution of
the homogenious Bethe--Salpeter equation \cite{2gam}:
\begin{equation}
\varGamma(p,P)=Z^{-1}g_S\left(1+\frac{\lambda_2}{\lambda_1}\frac{p(p-P)}{\beta^2}\right).
\label{vGZ}
\end{equation}
This vertex is to be normalised \cite{LS}, which gives for the scalar coupling the formula
\begin{equation}
\frac{g_S^2}{4\pi}=32m\sqrt{m\varepsilon}\left(1+2\frac{\lambda_2}{\lambda_1}\frac{m^2}{\beta^2}\right),
\label{gSdef}
\end{equation}
which coincides with the point-like result (\ref{gS0}) as $\beta\to\infty$. With the scalar vertex, the dressed kaon propagator, and the
dressed photon emission vertex in hand we are in a position to evaluate the width of the scalar two-photon decay up to the order $1/\beta^2$.
The amplitude of the process is given by the set of diagrams formally coinciding with those for the point-like vertex,
depicted at fig.\ref{3d}. Notice, however, an important difference: all ingredients are dressed now, and this is a necessary condition to
preserve gauge invariance beyond the point-like limit. The only quantity which should not be dressed is the $KK\gamma\gamma$ vertex in the 
third diagram. Indeed, the scalar vertex $\varGamma$ obeys the Bethe--Salpeter equation and thus absorbs all dressing diagrams. 

In view of
the fact that we deal with an explicitly gauge--invariant amplitude, we use the trick suggested in \cite{CIK} and, in order to extract the
amplitude, we read-off the coefficient at the structure $q_1^{\nu}q_2^{\mu}$ in the transition matrix element
\begin{equation}
iW=M(P^2)[q_1^{\nu}q_2^{\mu}-g^{\mu\nu}(q_1q_2)]
\epsilon_{1\mu}^*\epsilon_{2\nu}^*,\quad P=q_1+q_2,
\label{vertex}
\end{equation}
which is a particular case of (\ref{W1}) adapted for the two--photon case. Then
\begin{equation}
M(m_S^2)=M^{(0)}(m_S^2)+\frac{\lambda_2}{\lambda_1}\frac{m^2}{\beta^2}M^{(1)}(m_S^2),
\label{Mfull}
\end{equation}
and, by an explicit calculation, one can find that $M^{(1)}(m_S^2)=0$. Therefore, no large corrections of order $m^2/\beta^2$ appear for the
point-like result (\ref{beauty}).

We conclude therefore, that finite--range effects give only moderate
corrections to the point-like predictions (of order $10\div 20\%$ in the amplitude), provided they are included in a 
self-consistent and gauge-invariant way  \cite{Markushin,mol0,2gam}. We refer to the point-like results presented in tab.1 as to the molecule
model predictions for the radiative decays involving scalars.
For the sake of comparison, we quote in tab.\ref{t1} the results of calculations in the quark--antiquark assignment for the scalars, which
can be obtained with the help of the results of \cite{KR,qgamgam}. From tab.1 one can conclude that experimental data are well described in
the molecule assignment for the scalars (a recent result by Belle \cite{belle07} 
$\Gamma(f_0(980)\to\gamma \gamma)=0.205^{+0.095}_{-0.083}({\rm stat})^{+0.147}_{-0.117}({\rm syst})$ keV 
gives an even better coincidence with the point-like prediction). Furthermore, predictions for the radiative decays of scalars 
with massive vectors in the final state demonstrate a clear hierarchy, depending on the assignment prescribed to the scalar mesons. This
makes these decays an extremely promising tool in establishing the nature of the $a_0/f_0(980)$.

\section{Acknowledgements}

This work was supported by the Federal Agency for Atomic Energy of Russian 
Federation and by grants NSh-843.2006.2, DFG-436 RUS 113/820/0-1(R), 
RFFI-05-02-04012-NNIOa, and PTDC/FIS/70843/2006-Fi\-si\-ca.

\end{document}